\documentclass[aps,prl,showpacs,amsmath,twocolumn,amssymb,superscriptaddress,letterpaper]{revtex4}
\usepackage{times}
\usepackage{amsfonts}
\usepackage{mathrsfs}
\usepackage{graphicx}
\usepackage{dcolumn}
\usepackage{bm}
\usepackage{color}

\usepackage[colorlinks,bookmarks=false,citecolor=blue,linkcolor=red,urlcolor=blue]{hyperref}
\usepackage{bibunits}

\def\be{\begin{equation}}       \def\ee{\end{equation}}
\def\bea{\begin{eqnarray}}      \def\eea{\end{eqnarray}}

\begin{document}

\begin{bibunit}

\title{   i-Josephson Junction as Topological  Superconductor }
\author{Zhesen Yang}
\affiliation{Beijing National Laboratory for Condensed Matter Physics,
and Institute of Physics, Chinese Academy of Sciences, Beijing 100190, China}
\affiliation{University of Chinese Academy of Sciences, Beijing 100049, China}

\author{Shengshan Qin}
\affiliation{Beijing National Laboratory for Condensed Matter Physics,
and Institute of Physics, Chinese Academy of Sciences, Beijing 100190, China}
\affiliation{University of Chinese Academy of Sciences, Beijing 100049, China}

\author{Qiang Zhang}
\affiliation{Beijing National Laboratory for Condensed Matter Physics,
and Institute of Physics, Chinese Academy of Sciences, Beijing 100190, China}

\author{Chen Fang}
\affiliation{Beijing National Laboratory for Condensed Matter Physics,
and Institute of Physics, Chinese Academy of Sciences, Beijing 100190, China}

\author{Jiangping Hu}\email{jphu@iphy.ac.cn}
\affiliation{Beijing National Laboratory for Condensed Matter Physics,
and Institute of Physics, Chinese Academy of Sciences, Beijing 100190, China}
\affiliation{Kavli Institute of Theoretical Sciences, University of Chinese Academy of Sciences,
Beijing, 100190, China}
\affiliation{Collaborative Innovation Center of Quantum Matter,
Beijing, China}

\date{\today}
\begin{abstract}Superconducting states with broken time reversal symmetry are rarely found in nature. Here  
we predict that  it is inevitable that the  the time reversal symmetry is broken spontaneously in a superconducting Josephson junction formed by two superconductors with different pairing symmetries dubbed  as i-Josephson junction.  While the leading conventional Josephson coupling vanishes in such an i-Josephson junction,  the second order  coupling from tunneling always generates chiral superconductivity orders  with broken time reversal symmetry.  Josephson frequency in the i-junction is doubled, namely $\omega=4eV/h$. The result can not only provide a way to engineer topologically trivial or nontrivial time-reversal breaking superconducting states, but also be used to determine the pairing symmetry of unconventional superconductors. 
\end{abstract}


\maketitle
Introduction---Van der Waals (vdW) Josephson junction\cite{Machida2016}, which is contacted by two close-by superconducting (SC) layers by vdW forces, has been realized in layered dichalcogenide superconductors recently\cite{Machida2016, Hu-Jong Lee2017}. This provides a platform to investigate the properties of two SC layers with different pairing symmetries forming in the junction. In general, the physics of the junction is controlled by  the relative phase between the two SC order parameters, $\Delta\theta$.   In a conventional  Josephson junction which is formed by two s-wave  SC layers,   $\Delta\theta$  is typically zero in the absence of external or internal  magnetic fields. It can be  turned to nonzero  by magnetic fields that break the time reversal symmetry of the system explicitly.  However, in the unconventional Josephson junction, $\Delta\theta$ can be nonzero in the ground state without external nor internal magnetic fields\cite{Golubov2004, Goldobin2007, Goldobin2012}. A special case $\Delta\theta=\pm \pi/2$, which breaks time-reversal symmetry, is called chiral SC in the literature\cite{Maeno2003}.


Superconductors with spontaneously time-reversal symmetry breaking (TRB) pairing states\cite{Uemura1993,Koren2002,Kadono2003,Sigrist1998,Kapitulnik2015,Johrendt2013,Xia2017,Timm2017,Miguel2018}  have been widely sought.   The  most intriguing property of a TRB SC  is the nontrivial topology, namely, a TRB SC can be a topological superconductor (TSC)\cite{review1,review2,review3,review4}, e.g. topological $p+ip$\cite{TSC_Green,TSC_Kitaev,TSC_Das} and $d+id$\cite{Laughlin1998, Yao2013, Schaffer2012,Brydon2015,  Yonezawa2016, Honerkamp2014} superconductors. The former $p+ip$ TSC can be realized in many spin-orbital coupling systems\cite{TSC_Fu,TSC_Ando,TSC_Jia,TSC_Das,TSC_Oppen,TSC_Fisher,TSC_Mourik,TSC_Shtrik,TSC_Xu,TSC_Furdyna}, while the latter $d+id$ TSC has only been proposed in honeycomb lattice systems\cite{Hur2015}, such as doped graphene\cite{Jiang2008,Schaffer2012, Honerkamp2014, Schaffer2013, Baskaran2010}, single TiSe$_{2}$ layer\cite{Efremov2014} and bilayer Silicene\cite{Yao2013}. Although the $d+id$ TSC  exhibits many interesting phenomena, such as quantized boundary current\cite{Laughlin1998, Honerkamp2014}, spontaneous magnetization\cite{Laughlin1998, Golub2003},  quantized spin and thermal Hall conductance\cite{Golub2003, Honerkamp2014}, and geometric effects\cite{Lee2004}, there is no strong experimental evidence  to support the presence of this chiral SC state.

\begin{figure}[t]
\centerline{\includegraphics[height=4cm]{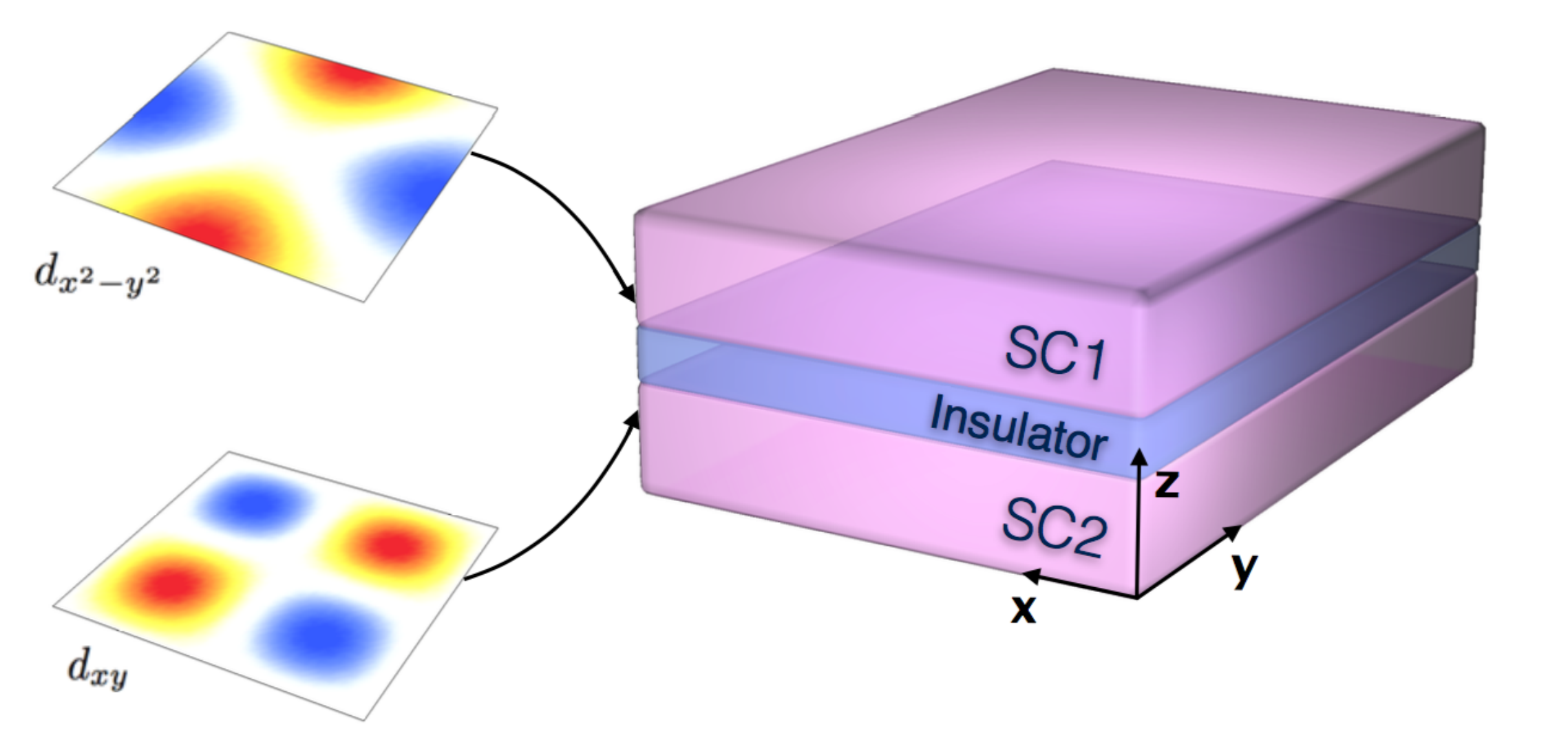}}
\caption{(color online). The sketch of a vdW Josephson junction with  two SC layers having two different pairing symmetries, e.g. $d_{x^2-y^2}$ and $d_{xy}$. }
\label{junction}
\end{figure}

Here,  we  ask whether a TRB SC can be spontaneously formed  in a vdW Josephson junction. We show that the TRB takes place spontaneously  in this Josephson junction formed by two SC layers with different pairing symmetries as illustrated in FIG. \ref{junction}. For example, a $d+id$ TRB SC can be engineered in a junction with a $d_{x^{2}-y^{2}}$  SC layer close to a $d_{xy}$  one. Furthermore, we prove that this $d+id$ SC constructed in this way is also a TSC. The junction has a distinct Josephson frequency, $\omega=4eV/h$, which is  twice of  the conventional Josephson frequency $2eV/h$.  We discuss possible experimental realizations for this type of junctions.  The results can not only help to realize novel SC states and design new SC qubit devices\cite{Makhlin2001, You2005}. 
but also be used to determine the pairing symmetry of an unknown SC by the unique feature of Josephson frequency.

Before we discuss specific models, we first present a general argument.  Considering a general Bogoliubov-De Gennes (BDG)  Hamiltonian  of  two SC layers  connected through tunneling and expanding  the free energy  up to  the fourth order of the tunneling,  the free energy can be generally written as\cite{Golubov2004, Goldobin2007, Goldobin2012, SM1}
\begin{equation}\mathcal{F}=\mathcal{F}_0-J \cos\Delta\theta+g \cos^{2}\Delta\theta,
\label{freeenergy}
\end{equation}
where the first term is the relative phase independent term, the second term is the conventional Josephson coupling term and the last term can lead to spontaneous TRB.  In a conventional Josephson junction, $J$ is positive and much larger than $g$ so that the third term can be ignored.  Here the main finding  is when two SC layers in a vdW Josephson junction have different pairing symmetries, $J$ vanishes and $g$ becomes the leading coupling from tunneling.  Remarkably, $g$ is always positive\cite{SM1}. Thus, to minimize the free energy, $\Delta\theta=\pm \pi/2$, which breaks TRS spontaneously. Such a vdW Josephson junction is called i-Josephson junction in this paper.

\begin{figure}[t]
\centerline{\includegraphics[height=7.5cm]{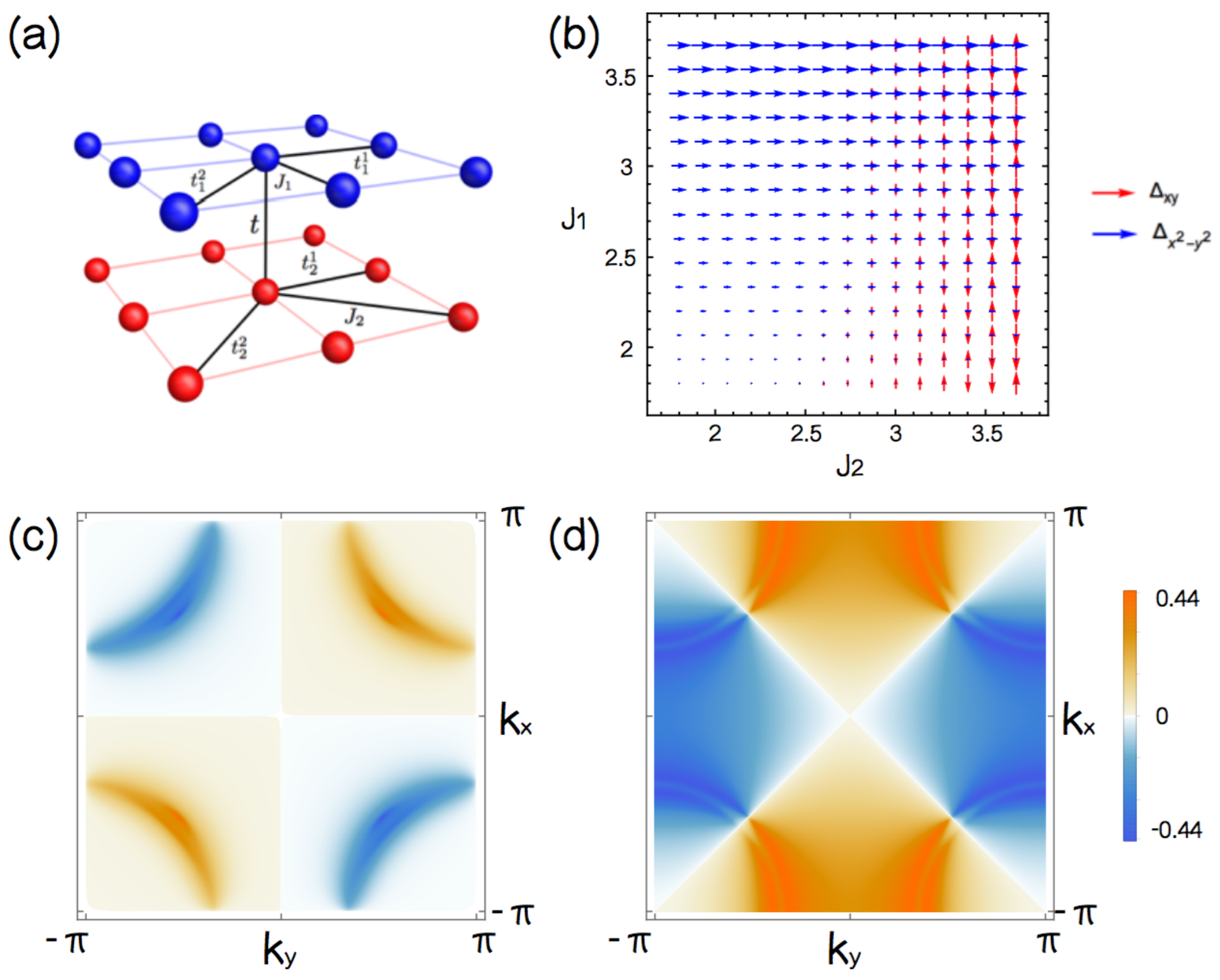}}
\caption{(color online). The sketch of the two layer model and  calculated SC orders from the mean field calculation:  (a) two layer $t-J$ models with the top layer $J_1$ and bottom layer $J_2$; (b) superconducting orders $\Delta_{xy}$ (red arrow) and $\Delta_{x^2-y^2}$ (blue arrow) as function of exchange strength $J_{1,2}$ with other parameters given in the main text. The  magnitudes and phases of pairing orders are represented by the lengths and directions of the arrows; (c) and (d) show the real and image parts of $d_{x^{2}-y^{2}}+id_{xy}$ order in the first layer with $J_{1}=3.4$ and $J_{2}=3$. The Fermi surfaces of the two layer Hamiltonian are represented by the red dashed lines, which coincide the arcs in BZ.
\label{P1}}
\end{figure}

{\em Model}---More specifically, the above analysis can be  modeled  by a two band superconductor in which the two bands have  different SC orders $\Delta_{\alpha}({\bf k})e^{i\theta_{\alpha}}$ and $\Delta_{\beta}({\bf k})e^{i\theta_{\beta}}$  where  $\theta_\alpha$ and $\theta_\beta$ are the SC phases. Their relative phase, $\Delta\theta=\theta_\beta- \theta_\alpha$, is a physical quantity when   the tunneling between two bands is induced.    The general BdG Hamiltonian  can be written as
\begin{equation}
\begin{aligned}
H&=\sum_{\bf k}\Psi^{\dag}_{\bf k}\mathcal{H}({\bf k})\Psi_{\bf k},\\
\end{aligned}
\label{eq22}
\end{equation}
where $\Psi_{{\bf k}}=(d_{{\bf k},\alpha,\uparrow}, d^{\dag}_{-{\bf k},\alpha,\downarrow}, d_{{\bf k},\beta,\uparrow}, d^{\dag}_{-{\bf k},\beta,\downarrow})^T$ and $\mathcal{H}({\bf k})$ is defined as
\begin{equation}
\begin{aligned}
\mathcal{H}({\bf k})&=
\begin{pmatrix}
\epsilon_{\alpha}({\bf k}) & \Delta_{\alpha}({\bf k})e^{i\theta_{\alpha}} & t({\bf k}) & 0 \\
\Delta_{\alpha}({\bf k})e^{-i\theta_{\alpha}} & -\epsilon_{\alpha}({\bf k}) & 0 & -t({\bf k})\\
t({\bf k}) & 0 & \epsilon_{\beta}({\bf k}) & \Delta_{\beta}({\bf k})e^{i\theta_{\beta}}\\
0 & -t({\bf k}) & \Delta_{\beta}({\bf k})e^{-i\theta_{\beta}} & -\epsilon_{\beta}({\bf k})\\
\end{pmatrix}.
\end{aligned}
\label{eq2}
\end{equation}
The tunneling $t({\bf k})$ is taken to be real, and the eigenvalues of this Hamiltonian for each $\bf k$ are
\begin{widetext}
\begin{equation}
E^{\pm\pm}=\pm\frac{1}{\sqrt{2}}\sqrt{\epsilon^{2}_{\alpha}+\epsilon^{2}_{\beta}+\Delta_{\alpha}^{2}+\Delta_{\beta}^{2}+2t^{2}\pm\sqrt{(\epsilon^{2}_{\alpha}-\epsilon^{2}_{\beta}+\Delta_{\alpha}^{2}-\Delta_{\beta}^{2})^{2}+4t^{2}[(\epsilon_{\alpha}+\epsilon_{\beta})^{2}+|\Delta_{\alpha}e^{i\theta_{\alpha}}-\Delta_{\beta}e^{i\theta_{\beta}}|^{2}]}}.
\end{equation}
\end{widetext}

At zero temperature, the  free energy is  $\mathcal{F}=\sum_{{\bf k}}(E^{-+}({\bf k})+E^{--}({\bf k}))$.   We can expand the free energy  up to the fourth order of $t({\bf k})$. The free energy is given by Eq. (\ref{freeenergy}), in which the parameters  can be specified as
\begin{eqnarray}
& & J= \sum_{\bf k}[t^2({\bf k})g_{1}({\bf k})+t^4({\bf k})g_3({\bf k})]\Delta_{\alpha}({\bf k}) \Delta_{\beta}({\bf k}) \\
& & g= \sum_{\bf k} t^4({\bf k})g_2({\bf k})\Delta^2_{\alpha}({\bf k})\Delta^2_{\beta}({\bf k}).
\end{eqnarray}
The explicitly form of $g_i({\bf k})$ are shown in the Supplementary Materials. While the functions of $g_{i}({\bf k})$ are very lengthy, we can analyze their symmetry characters. For convenience, we consider a square lattice symmetry classified by the $C_{4v}$ point  group. One can notice all the parameter functions except $\Delta_{i}({\bf k})$ belong to the $A_{1}$ irreducible representation of $C_{4v}$. Thus,  if $\Delta_{\alpha}({\bf k})$ and $\Delta_{\beta}({\bf k})$  belong to different irreducible representations, namely, they have  different pairing symmetries,  the conventional Josephson coupling, $J$, vanishes because of  the symmetry constraint. Therefore, the ground state is determined by the sign of $g$, that is if $g>0$, $\Delta\theta=\pm\pi/2$ and if $g<0$, $\Delta\theta=0,\pm\pi$. In the Supplementary Materials, we have shown the positive natural of $g_2({\bf k})$ for all ${\bf k}$ in the BZ. Thus the relative phase in the ground state is always $\pm\pi/2$. 


{\em Mean filed calculation}---The above results can be further examined in a specific model. We consider two layered superconductors to obtain the TRB $d+id$ order. The $d$-wave SC state develops  naturally  if the SC pairing is driven by local antiferromagnetic fluctuations\cite{scalapino, anderson}. Theoretically, we can use the $t-J$ model to model the $d$-wave SC state.  Following the well-known result, we consider the following junction as illustrated in FIG. \ref{P1} (a). The pairing interactions on each layer are  attributed to  anti-ferromagnetic exchange interactions $J_\alpha$.   We only consider the  nearest neighbor (NN) exchange $J_1$, and  the next nearest neighbor (NNN) $J_2$ in the top and bottom layers respectively. Including the tunneling coupling between the two layers, the overall Hamiltonian can be written as $H=H_{J_{1}}+H_{J_{2}}+H_{t}$, where
\begin{equation}\begin{aligned}
H_{J_{\alpha}}=\sum_{{\bf k},\sigma}\xi&_{\alpha}({\bf k})d^{\alpha\dag}_{{\bf k},\sigma}d^{\alpha}_{{\bf k},\sigma}+\sum_{i,\delta=\hat{x},\hat{y}}J_{\alpha}(\vec{S}^{\alpha}_{i}\cdot\vec{S}^{\alpha}_{i+\delta}-\frac{1}{4}n^{\alpha}_{i}n^{\alpha}_{i+\delta}),\\
&H_{t}=t\sum_{\alpha,{\bf k},\sigma}d^{1\dag}_{{\bf k},\sigma}d^{2}_{{\bf k},\sigma}+h.c..
\label{MF}
\end{aligned}\end{equation}
Here the tunneling term $t$ is chosen to be real and independent of ${\bf k}$ for simplicity. We also  drop the double occupancy projection operators which is required in the standard $t-J$ model because the double occupancy projection in the mean field level can be treated as an overall renormalization factor to the band dispersion\cite{Liu1988,Hu2008}.  Therefore, it does not affect the qualitative result.  The sketch of this model is shown in FIG. \ref{P1} (a).

In the mean field solution, we can compare the energies of the $s$-wave and $d$-wave SC states.  The self-consistent mean field solutions for the d-wave SC states are given by\cite{SM2}
\begin{equation}
\frac{\Delta_{x^{2}-y^{2}}({\bf k})}{\cos k_{x}-\cos k_{y}}=\sum_{{\bf k}'}-\frac{2J_{1}}{N}(\cos k'_{x}\pm\cos k'_{y})\langle d^{1}_{-{\bf k}',\downarrow}d^{1}_{{\bf k}',\uparrow}\rangle
\end{equation}
\begin{equation}
\frac{\Delta_{xy}({\bf k})}{\sin k_{x}\sin k_{y}}=\sum_{{\bf k}'}-\frac{8J_{2}}{N}\sin k'_{x}\sin k'_{y}\langle d^{2}_{-{\bf k}',\downarrow}d^{2}_{{\bf k}',\uparrow}\rangle.
\end{equation}
We take the band dispersion,
 \begin{eqnarray}\xi_{\alpha}=-2t^{1}_{\alpha}(\cos k_{x}+\cos k_{y})-4t^{2}_{\alpha}\cos k_{x}\cos k_{y}+\mu_{\alpha}-\mu, \end{eqnarray}
where $t^{1(2)}$ indicates the NN (NNN) hopping and $\alpha=1,2$, corresponding the top and bottom layers. $\mu_\alpha$ is the corresponding on-site energy in each layer and $\mu$ is the chemical potential.  Without the tunneling, in the mean field solution shown in the Supplementary Materials,  we find  that  the $d_{x^{2}-y^{2}}$-wave and $d_{xy}$-wave  orders are favored on the top and bottom layers respectively  when the parameters are set as $t^{1}_{1}=0.88$, $t^{2}_{1}=-0.35$, $t^{1}_{2}=1.67$ , $t^{2}_{2}=-0.33$,  $\mu_{1}=-0.4$, $\mu_{2}=-1.2$  and $\mu=0$.

Turning on the layer  tunneling  and taking $t=0.4$, the phase diagram is plotted in the Fig. \ref{P1} (b) as the function of $J_{1,2}$. The lengths of the vectors in Fig. \ref{P1} (b) represent the strength  of the orders and the directions relate to the phases. As $J_{1}$ and $J_{2}$ increase, both $d$-wave orders become stronger and the relative phase  maintains to be $\pm\pi/2$.    The imaginary and real parts of the order parameter  in the first layer are shown in Fig. \ref{P1} (c) and (d). Clearly, the real part has $d_{x^2-y^2}$ symmetry and the imaginary part has  $d_{xy}$ symmetry.  The $d_{xy}$ order in the first layer is induced through  the proximity effect from the second layer.  This result  is consistent with  our previous analysis.

\begin{figure}[t]
\centerline{\includegraphics[height=8cm]{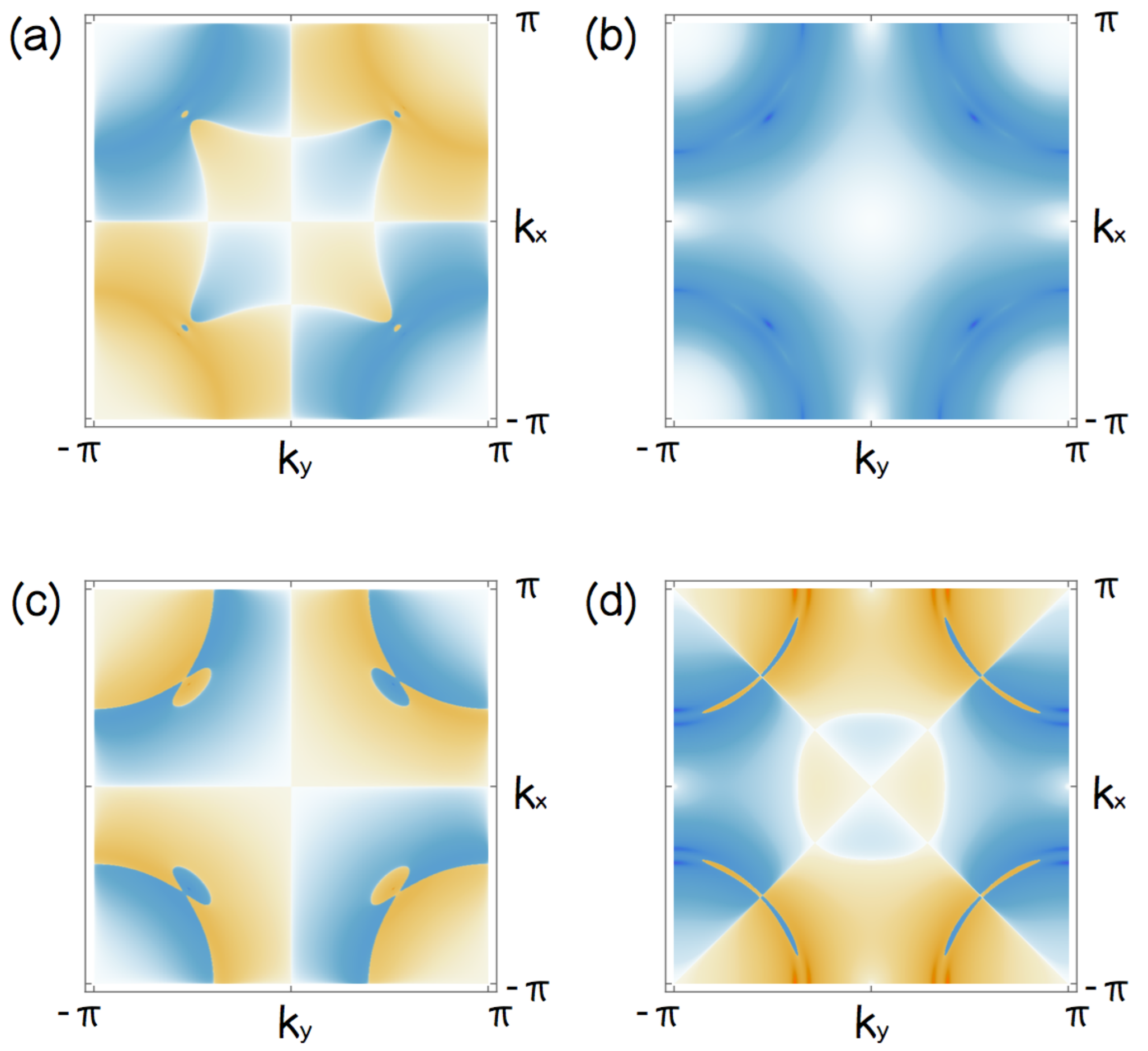}}
\caption{(color online). The Berry curvature for different TRB orders in the Hamiltonian (\ref{eq2})  with  parameters $t^{1}_{1}=0.88$, $t^{2}_{1}=-0.35$, $t^{1}_{2}=1.67$ , $t^{2}_{2}=-0.33$,  $\mu_{1}=-0.4$, $\mu_{2}=-1.2$, $\mu=0$, $t=0.4$, $\Delta_\alpha=0.8183$, $\Delta_{\beta}=0.5562$, $\theta_{\alpha}=0$ and $\theta_{\beta}=\pi/2$. (a), (b), (c) and (d) represent $d+is'$, $d+id'$, $d+is$ and $d'+is$ orders, respectively. According to our analysis, the Berry field of $d\pm is(s')$ belong to $B_2$ IR, $d'\pm is(s')$ belong to $B_1$ IR and only $d+id'$ belongs to $A_1$ IR. Only $d\pm id^\prime$ in (b) shows a nonzero Chern number. }
\label{P2}
\end{figure}

{{\em Topological analysis}---Now we discuss the topological properties of the above spontaneous TRB SCs.  The state in the i-junction can be topologically nontrivial for a  $d\pm id$ state but  trivial  for an $s\pm id$ state.  To show this, we can analyze  the symmetry property of the Berry curvature.
Starting from the Hamiltonian (\ref{eq2}) with $\theta_\alpha=0$ and $\theta_\beta=\pi/2$,  the band dispersions $\epsilon_{\alpha}(\bf{k})$ and $\epsilon_\beta(\bf{k})$ both belong to the $A_1$ irreducible representation (IR) of $C_{4v}$.  We consider the  $\sigma_{v}$ symmetry operation which maps  ${\bf k}=(k_x,k_y)\rightarrow\tilde{\bf k}=(-k_x, k_y)$ or $(k_x,-k_y)$. In the  $s\pm id$ state, if the tunneling term $t(\bf{k})$ belongs to $A_1$ or $B_1$ IR,
the Hamiltonian is invariant under $\sigma_v$ operation.  If the tunneling term $t({\bf{k}})$ belongs to $B_2$ IR,  under a $\sigma_v$ operation, the Hamiltonian  becomes $\tilde{\mathcal{H}}(\tilde{{\bf{k}}})$, which can be expressed as $\tilde{\mathcal{H}}(\tilde{{\bf{k}}})=\tau_z\mathcal{H}({\bf{k}})\tau_z$. The corresponding eigenstate  becomes  $|\tilde{u}_n(\tilde{{\bf{k}}})\rangle=\tau_z|u_n({\bf{k}})\rangle$.  In both cases,
considering the definition of Berry curvature
\begin{equation}
B({\bf{k}})=i\sum_{n\in occ}\epsilon_{k_xk_y}\langle\partial_{k_x} u_n({\bf{k}})|\partial_{k_y} u_n(\bf{k})\rangle,
\label{eq7}
\end{equation}
one can find that the Berry curvature  changes sign under $\sigma_v$ operation so that the total Chern number is zero.
But for the  $d\pm id'$ state, the Berry curvature is invariant under $C_4$, $\sigma_v$ and $\sigma_d$ operation. This means  that the nonzero Chern number is not forbidden by any symmetry operations.  Thus, under some suitable parameters, the system can be a topological $d\pm id$ SC. This is clearly shown in FIG. \ref{P2}.

The above conclusion can be  numerically verified.  We perform numerical calculation for the model in Eq. (\ref{MF}). We calculate the topologically protected edge states  in a stripe lattice as shown  in FIG. \ref{P3}. The parameters in the calculation are set to be $J_{1}=3.4$, $J_{2}=3$ in Eq. (\ref{MF}), and the corresponding mean field SC orders strength are $\Delta_{x^{2}-y^{2}}=0.8183$, $\Delta_{xy}=0.5562i$ and $\Delta_{x^2+y^2}=\Delta_{x^2y^2}=0$.  Under these parameters, there are four chiral modes on each edge, which are corresponding to  a Chern number equal to minus four, as shown in FIG. \ref{P3}.

{\em Experimental signatures}---There are three smoking-gun signatures for the above topological $d+id$ i-Josephson junction.  The first one is the topologically protected  edge state as shown in FIG. \ref{P3}. On the edge, one can use superconducting quantum interference  microscopies to detect spontaneously generated supercurrents\cite{Laughlin1998,Honerkamp2014,Kirtley2007}.  The second one is that the Josephson frequency  doubles  the conventional one. For a conventional Josephson junction, the Josephson frequency is given by $\omega_0=2eV_0/h$, where $V_0$ is the applied external voltage on the junction. For the i-junction,  the modified Josephson equations are
\begin{equation}
\begin{aligned}
&I=I_0\sin 2\Delta\theta,\\
&\frac{d \Delta\theta}{dt}=-\frac{2e}{\hbar}V_0.
\label{E8}
\end{aligned}
\end{equation}
The AC Josephson current is $I=I_0\sin(2\Delta\theta_0-4eV_0t/\hbar) $. The corresponding Josephson frequency is $\omega_i=2\omega_0=4eV_0/h$, which is twice of the ordinary Josephson frequency.

\begin{figure}[t]
\centerline{\includegraphics[height=7cm]{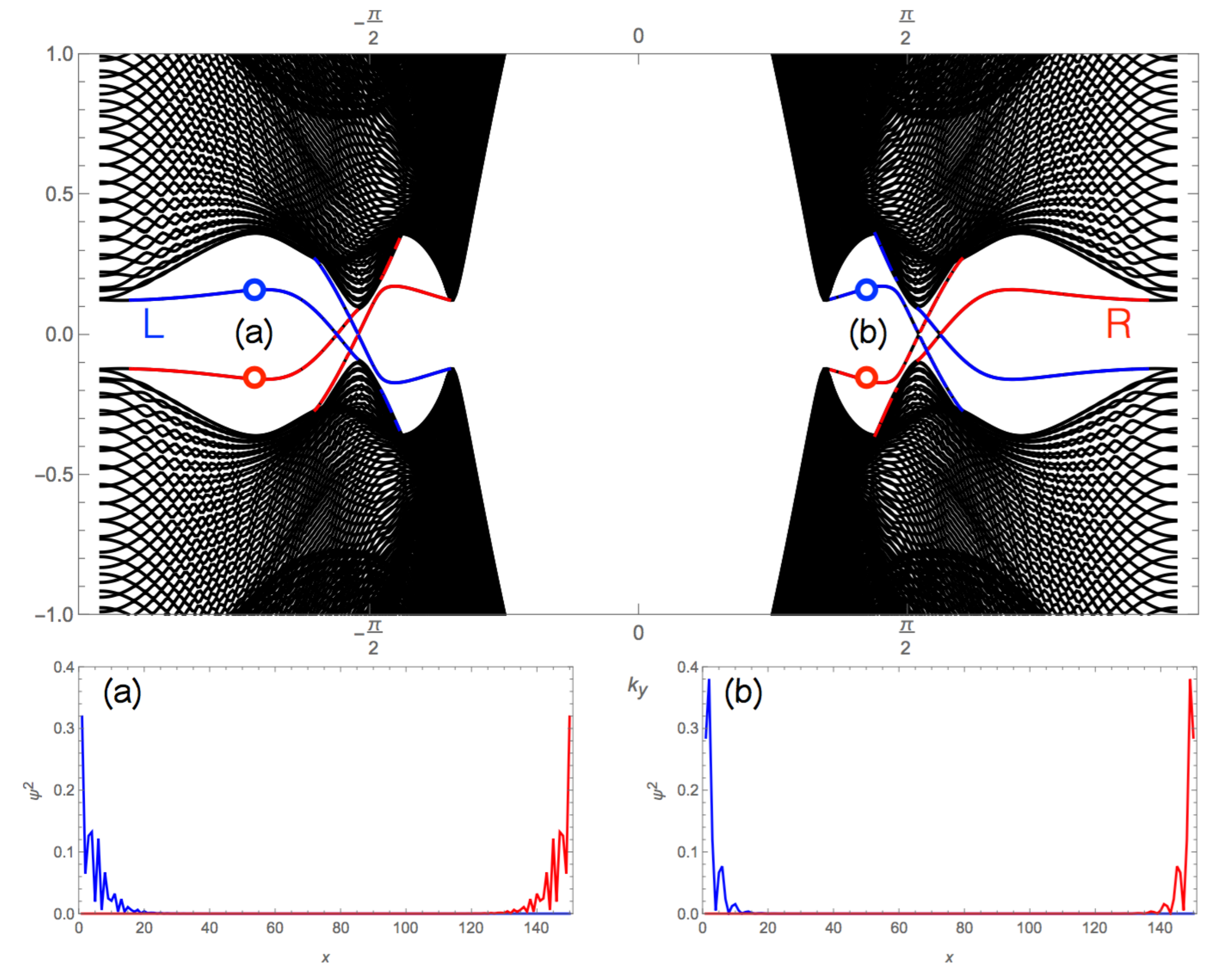}}
\caption{(color online). Edge states of $d+id'$ superconductivity in the Hamiltonian (6), the parameter we choose is $J_{1}=3.4$, $J_{2}=3$, and the corresponding self-consistent mean field SC orders are $\Delta_{x^{2}-y^{2}}=0.8183$, $\Delta_{xy}=0.5562i$. On the top panel is the band structure with left edge state blue and right red. The bottom panel are the corresponding distribution of edge states. }
\label{P3}
\end{figure}

The third experimental signature is the magnetic field dependence of the critical current.  When a magnetic field $B$ is applied to a  conventional Josephson junction with a length L and the penetration depth $W$,  the critical current is
\begin{eqnarray}
I_c=j_0\left|\frac{\sin(\pi\Phi/\Phi_0)}{\pi\Phi/\Phi_0}\right|,
\end{eqnarray}
where  $\Phi_0=h/2e$, is the flux quantum and $\Phi=BWL$.
In the i-Josephson junction,  it is easy to show that
\begin{eqnarray}
I_c=j_0\left|\frac{\sin(2\pi\Phi/\Phi_0)}{2\pi\Phi/\Phi_0}\right|.
\label{E9}
\end{eqnarray}
The oscillation pattern is changed. Notice the last two experimental signatures are valid for all i-Josephson junctions.

{\em Engineering TSC}---Previously, the TSCs have already been  proposed in  $p$-wave superconductors\cite{TSC_Kitaev}, TI surface states\cite{TSC_Fu,TSC_Ando,TSC_Jia} and semiconductor nanowires\cite{TSC_Das,TSC_Oppen,TSC_Fisher,TSC_Mourik,TSC_Shtrik,TSC_Xu,TSC_Furdyna}.   These proposals all focus on the $0d$ Majorana bound states, not the $1d$ Majorana chiral edge states. Recently, the chiral Majorana modes have been observed in the  quantum anomalous Hall insulator-superconductor structure\cite{Zhang}. Compared to the previous work, the major advantage here  is that TSC and the corresponding chiral edge state can be realized with conventional d-wave superconductors, such as cuprates, and  no external magnetic field\cite{Sau2010} or topological non-trivial band structures are needed.  Thus, in principle, our method allows TSC to operate at very high temperature because of the high SC transition temperature of cuprates. 

Recently, the advances in vdW heterostructure technology provide an effective way to engineer the rotation angle between the two SC layers with a high accuracy\cite{Tutuc2016, Cao2018}. Furthermore, the vdW junction is defect-free contacted and has a strong proximity coupling\cite{Machida2016, Hu-Jong Lee2017}, which renders a larger value of $g$ in Eq. (\ref{freeenergy}). Thus an explicit design of aTSC i-junction is to  align two identical d-wave superconductors along the z direction with a relative $\pi/4$ in-plane angle\cite{angle}.  This design can be implemented  by  recent rapid technological progress in engineering heterostructures.


{\em Engineering TRB state}---The result also allows us to engineer new exotic SC states with TRB, e.g. $s+id$ pairing state. Superconductors with these type of pairing states have been widely searched. However, success has been very limited. So far, spontaneous TRB in SC states have been rarely observed.

The above physics can also be potentially realized in bulk materials. For example, it has been theoretically suggested that  the FeAs layer, the  building block in iron-based superconductors, and the CuO$_2$ layer, the building block in cuprates, can be hybridized to form a hybrid crystal\cite{daix}.   Following our results, in such a hybrid crystal, the time reversal symmetry must be broken as FeAs\cite{Hu2008,fang} and CuO$_2$\cite{scalapino, anderson}  are known to favor $s$-wave and $d$-wave pairing symmetries respectively.  The superconducting state in such a material must be $s\pm i d$.

{\em Applications}--- The i-Josephson junction can be used to determine the pairing symmetry of an unknown superconductor. This is based on the fact that the Josephson frequency will be doubled if two SC layers have different pairing symmetries.  The i-Josephson junction can also be used to make a quantum qubit because  the free energy    has two minima.  It becomes a natural two-level system to  form a qubit.   Other excited states have much higher energy so that the two-level system is well protected.

In summary,  we have shown that  the i-Josephson junction is an inevitable result when a Josephson  junction is formed by two superconductors with different pairing symmetries. The Josephson frequency is doubled. A TSC with a $d\pm id$ pairing symmetry can be achieved in this way. The result also provides a method to design TRB $s+id$ superconductor.

\begin{acknowledgments}
{\em Acknowledgement.---}We thank useful discussions with X.X. Wu and Yong P. Chen. The work is supported by the Ministry of Science and Technology of China 973 program (No. 2015CB921300, No. 2017YFA0303100), National Science Foundation of China (Grant No. NSFC-1190020, 11534014, 11334012), the Strategic Priority Research Program of CAS (Grant No.XDB07000000), and the Key Research Program of the CAS(Grant No. XDPB08-1).

\end{acknowledgments}

}
\end{bibunit}

\end{document}